\documentclass[pra,twocolumn,amsmath,amssymb,superscriptaddress,showpacs]{revtex4}
\usepackage{graphics}
\usepackage{epsfig}

\newcommand{\ket}[1]{\ensuremath{\left|{#1}\right\rangle}}
\newcommand{\ad}{\ensuremath{a^\dagger}}
\def\be{\begin{equation}}
\def\ee{\end{equation}}
\def\eea{\end{eqnarray}}
\def\bea{\begin{eqnarray}}
\newcommand{\bd}{\ensuremath{b^\dagger}}
\newcommand{\va}[1]{\ensuremath{(\Delta#1)^2}}
\newcommand{\varho}[1]{\ensuremath{(\Delta_\rho #1)^2}}
\newcommand{\ex}[1]{\ensuremath{\left\langle{#1}\right\rangle}}
\newcommand{\exrho}[1]{\ensuremath{\left\langle{#1}\right\rangle}_{\rho}}

\begin{document}
\title{Entanglement Detection Based on Interference and Particle Counting}
\date{\today}
\begin{abstract}
A sufficient condition for entanglement in two-mode continuous
systems is constructed  with the interference visibility and
the uncertainty of the total particle number. 
The observables to be measured  
(particle numbers and particle number variances) 
are relatively easily accessible experimentally. 
The method may be used to detect entanglement in   
light fields or in Bose-Einstein condensates.
In contrast to the standard approach based on
entanglement witnesses, the condition is expressed in terms
of an inequality which is nonlinear in expectation values.
The condition is constructed using uncertainty relations with the particle number
and the destruction operators.
\end{abstract}
\author{G\'eza T\'oth}
\affiliation{Max-Planck-Institut f\"ur Quantenoptik,
  Hans-Kopfermann-Str. 1, Garching, D-85748, Germany.}  
\author{Christoph Simon}
\affiliation{Department of Physics, University of Oxford, Oxford
OX1 3PU, United Kingdom}
\author{Juan Ignacio Cirac}
\affiliation{Max-Planck-Institut f\"ur Quantenoptik,
  Hans-Kopfermann-Str. 1, Garching, D-85748, Germany.}  
\maketitle

\section {Introduction}

In spite of considerable efforts, the
separability of general mixed quantum states is still an open problem,
even if the whole density matrix is known.
The positivity of the partial transpose \cite{AP96} of the density matrix
is a necessary condition for separability; however, it is
a sufficient condition only for the $2 \times 2$ (two qubits) and 
$2 \times 3$ dimensional cases and for two-mode Gaussian states in continuous variable systems.
For higher dimensions there exist entangled states \cite{PH97} with positive
partial transpose and the separability problem is not fully solved. 
For bipartite low-rank density matrices an operational
criterion for separability is presented in Ref. \cite{HL00}.
A necessary and sufficient condition for entanglement of all bipartite
Gaussian continuous states is also known \cite{GK01}.
Entanglement can also be detected by several other methods, e.g., 
through witness operators 
whose expectation value 
is negative only for (some) entangled states \cite{HH96}.

In an experiment the density matrix is usually not known, 
only partial information is available on
the quantum state. One can typically measure a few observables
and still would like to detect some of the entangled states \cite{GH02}.
Thus to find a criterion for entanglement 
with easily measurable observables is crucial
for entanglement detection.
There are only few such criteria in the literature
\cite{SD01,DG00,RS00,HT02,CS03} 
for detecting entanglement in complicated situations
such as in many-particle or continuous variable systems.
The method described in Ref. \cite{SD01} 
detects entanglement
among cold atoms having two internal degrees of freedom
based on inequalities with the 
total angular momentum components. 
Ref. \cite{CS03}  derives a criterion for the entanglement between
two pairs of bosonic modes 
in terms of the total angular momentum and the particle
number. This criterion detects entangled states
that are close to singlet states of two large spins. 

A criterion for detecting entanglement between two modes
is given in \cite{DG00,RS00}. Ref. \cite{DG00} presents a scenario where
one just
has to measure the second moments of $x$ and $p$ for both systems. For
example, if the inequality
\be
(\Delta (x_A+x_B))^2 + (\Delta (p_A-p_B))^2 <2
\label{crit1}
\ee
is fulfilled, then the state is entangled. Here $(\Delta X)^2=
\langle X^2\rangle - \langle X\rangle^2$, 
and $x_k$ and $p_k$ are canonical
operators satisfying $[x_k,p_k]=i$. 
Note that the left hand side of Ineq. (\ref{crit1}) 
is quadratic in expectation values. However,
by setting $\ex{x_k}$ and $\ex{p_k}$
to zero by single-party unitary operations, 
only terms linear in expectation values remain
and the criterion is equivalent to an entanglement witness.

A generalization of criterion (\ref{crit1}) 
detects all entangled two-mode Gaussian states \cite{DG00,RS00}.
However, in many experimental situations non-Gaussian states are
prepared. For example, if one has $N$ photons and sends them
through a beam splitter or if one has $N$ atoms in some internal
state and applies an appropriate laser pulse, the state will be
\be
|\Psi\rangle= \frac{1} {\sqrt{2^N N!}} (a^\dagger+b^\dagger)^N |0,0\rangle
= \frac{1} {\sqrt{2^N}} \sum_{n=0}^N \sqrt{\binom{N}{n}}|n,N-n\rangle
\label{PPS}
\ee
Here $a$ and $b$ are annihilation operators which are defined
according to $x_A=(a+a^\dagger)/\sqrt{2}$. This entangled state is not
detected by the previous criterion as it will be demonstrated later.

In this paper, we will present a criterion which: (i) requires measuring
quantities which are easily accessible experimentally and; (ii)
detects entangled states in the vicinity of state (\ref{PPS}). 
The paper is organized as follows. 
In Sec. II. the entanglement criterion is derived.
In Sec III. it is discussed what states are detected by this
criterion and 
issues concerning its possible experimental applications
are also considered. A summary is given in Sec. IV.

\section {Entanglement criterion}

In this section we will show that for all separable states,
i.e. states that can be written as
\begin{equation}
\rho = \sum_k p_k \rho_k^A \otimes \rho_k^B,
\label{sep}
\end{equation}
the following expression
involving the variances of the total particle number $N:=\ad a +\bd b$
and of the operator $(a-b)$ is bounded from below as
\begin{equation}
\bigg\{ (\Delta_\rho N)^2+1\bigg\}
\bigg\{ (\Delta_\rho (a-b))^2+1\bigg\}
\ge \frac{\exrho{N}} {4} + \frac{1}{8},
\label{vnvbregion}
\end{equation}
where $(\Delta_\rho A)^2:= \langle A^\dagger A\rangle_\rho -
|\exrho{A}|^2$ (note that $A$ need not be Hermitian).

The physical motivation for this criterion comes from the observation, made
in the context of Bose-Einstein condensates in Ref. \cite{CS01} (also
explained in Sec. III), that it is
not possible to have a fixed total particle number --- corresponding to 
$(\Delta_\rho N)^2=0$ ---
and perfect interference --- corresponding to $(\Delta_\rho (a-b))^2=0$ --- 
at the same time, unless the system under 
consideration is in a highly non-classical, i.e. entangled, state.

In Sec. II.A a simple separability criterion will be proved.
In Sec. II.B this criterion
will be generalized. Technical details are in Appendix A.
Sec. II.C proves our main result, Ineq. (\ref{vnvbregion}).

\subsection{Simple criterion}

In this subsection a simple separability criterion will be derived  based
on uncertainty relations for the two subsystems. 
In order to understand the connection between the uncertainty 
relations and the necessary condition for separability, 
it is instructive first to review how the condition (\ref{crit1})
was obtained 
starting from the single-subsystem uncertainty relations \cite{DG00,HT02}
\be
\varho{x_{A/B}}+\varho{p_{A/B}}\ge 1. 
\label{uncxp}
\ee
For a separable state of the form (\ref{sep}) the sum of uncertainties of
the EPR type operators $x_A+x_B$ and $p_A-p_B$ can be written as
\bea
&&(\Delta_\rho(x_A+x_B))^2+\varho{(p_A-p_B)}\nonumber\\
&=&\sum_k p_k \bigg\{(\Delta_{\rho^A_k}x_A)^2 
+(\Delta_{\rho^A_k}p_A)^2\nonumber\\
&+&(\Delta_{\rho^B_k}x_B)^2+(\Delta_{\rho^B_k}p_B)^2\bigg\}\nonumber\\
&+&\sum_k p_k \bigg\{ \ex{(x_A+x_B)}_{\rho^A_k\otimes\rho^B_k}^2-
\exrho{(x_A+x_B)}^2\nonumber\\   
&+&\ex{(p_A-p_B)}_{\rho^A_k\otimes\rho^B_k}^2-\exrho{(p_A-p_B)}^2\bigg\} \ge 2.
\label{xpineq}
\eea
In the equality it was exploited, that
for a product state the uncertainty of an EPR type 
operator splits into the sum of the corresponding single system uncertainties
as $(\Delta_{\rho^A_k\otimes\rho^B_k} (x_A+x_B))^2=
(\Delta_{\rho^A_k} x_A)^2+(\Delta_{\rho^B_k} x_B)^2$.
Based on the uncertainty relations 
(\ref{uncxp}) for the individual subsystems, A and B,
the first sum in Ineq. (\ref{xpineq}) is bounded from below by $2$.
Since the second sum is non-negative,
the left hand side of Ineq. (\ref{xpineq}) is also bounded by $2$ which
finishes the proof of criterion (\ref{crit1}).
Thus for separable states the sum of the variances 
of $x_A+x_B$ and $p_A-p_B$ has the same lower bound
as the sum of the corresponding single system uncertainties.
Note that this simple relationship holds because
the right hand side of the the uncertainty relation (\ref{uncxp})
is a {\it constant}.
On the other hand, this bound is not valid 
for non-separable states. In this case the sum of the
two uncertainties can even be zero \cite{DG00} since
$x_A+x_B$ and $p_A-p_B$ commute.

After reviewing the previous example with the EPR operators, 
we will prove that
all separable states fulfill
\begin{equation}
\Sigma_\rho:= (\Delta_\rho N)^2 + (\Delta_\rho (a-b))^2 \ge
f(\exrho{N}),
\label{IneqNb}
\end{equation}
where
\be
 f(N)=\sqrt{N+\frac{3}{4}}+\frac{\sqrt{3}}{2}-2.
\ee
This will be necessary later to obtain our main result,
Ineq. (\ref{vnvbregion}). Here $(\Delta_\rho N)^2$ is the
variance of the total particle number in the two-mode system
(i.e., in a two-mode electromagnetic field or 
Bose-Einstein condensate in a double-well potential) while 
$(\Delta_\rho (a-b))^2$ is related to the
variance of the phase difference between the two modes.

The proof of (\ref{IneqNb}) is the following.
For a separable state of the form (\ref{sep}), 
the sums of the two uncertainties in (\ref{IneqNb}) can be written as
$\Sigma_\rho=\Sigma_{\rho,0} + \Sigma_{\rho,1}$ where
\bea
\Sigma_{\rho,0} &=& 
\sum_k p_k \bigg\{(\Delta_{\rho_k^A} N_A)^2+(\Delta_{\rho_k^A}
a)^2\nonumber\\\label{summarho0}
&+&(\Delta_{\rho_k^B} N_B)^2+(\Delta_{\rho_k^B} b)^2
+ N_k^2 - \exrho{N}^2\bigg\},\\
\Sigma_{\rho,1} &=& \sum_k p_k |\langle (a-b) \rangle_{\rho_k}|^2-|\langle
(a-b)\rangle_\rho|^2.
\eea
Here $N_A:=\ad a$, $N_B:=\bd b$ and
$N_k:=\langle a^\dagger a+b^\dagger b \rangle_{\rho_k}$.
Using the Cauchy-Schwarz inequality, one
can show that $\Sigma_{\rho,1}\ge 0$, 
and thus $\Sigma_\rho \ge \Sigma_{\rho,0}$.

Next, we will need the following uncertainty relation 
proved in Appendix A.1-2
\be
R_\rho:=(\Delta_{\rho} N_A)^2+(\Delta_{\rho} a)^2 \ge L(\exrho{N_A}),
\label{Rrho}
\ee
where
\be
L(N)=\sqrt{N+\frac{3}{4}}-1.
\label{LN}
\ee
Obviously, the same inequality is true for subsystem $B$.
Ineq. (\ref{Rrho}) is an alternative of the number-phase uncertainty
without the problem of defining
an appropriate phase operator and the difficulties due to the $2\pi$ 
periodic nature of the phase \cite{L95}.

In our case 
the bound in the uncertainty relation (\ref{Rrho}) is not a constant,
but a {\it function} 
of an operator expectation value.
Thus the method presented for the EPR type operators cannot be used, 
and careful analysis of the different properties
of the function $L(N)$ must be done \cite{TermsKept}. 
Using inequality (\ref{Rrho}) and the fact that the bound (\ref{LN}) 
fulfills 
$L(N_1)+L(N_2)\ge L(N_1+N_2)+L(0)$ we obtain
\be
\Sigma_\rho \ge \sum_k p_k \bigg\{L(N_k)+L(0)+N_k^2 - \exrho{N}^2\bigg\}.
\label{last}
\ee
Now using the fact that $L(N_k)+N_k^2$ is a concave
function of $N_k$, we obtain $\Sigma_\rho \ge L(N)+L(0)$ which proves 
Ineq. (\ref{IneqNb}). 

Condition (\ref{IneqNb}) corresponds to a line on the 
$\varho{N}$ -- $\varho{(a-b)}$ plane
(solid line in the inset of Fig. \ref{fig_region}).
All separable states belong to points above
this line and fulfill Ineq. (\ref{IneqNb}).
Points below this line correspond to entangled states only.

\subsection{Generalization}

We would like to find more entangled states in the
$\varho{N}$ -- $\varho{(a-b)}$ plane.
In order to do that we generalize (\ref{IneqNb}) as
\begin{equation}
\Sigma_{\rho,w}:= w(\Delta_\rho N)^2 + (1-w)(\Delta_\rho (a-b))^2 \ge
f_w(\exrho{N})
\label{Sigmaw}
\end{equation}
where $0 < w < 1$ determines the relative weights of the two terms
and $f_w(N)$ is defined at the end of this subsection.
Ineq. (\ref{Sigmaw}) correspond a 
region above a line with slope $\frac{w}{1-w}$
(dashed lines in the inset of Fig. \ref{fig_region}).
These lines are the tangentials
of the curve enclosing all separable states.
The points below this curve correspond all to entangled states.

In order to obtain the lower bound $f_w(\exrho{N})$, we have to follow a
procedure similar to what was presented in the previous subsection.
For a separable state one obtains
\bea
\Sigma_{\rho,w}&\ge& \sum_k p_k\bigg\{w(\Delta_{\rho_k^A} N_A)^2
+(1-w) (\Delta_{\rho_k^A} a)^2\nonumber\\
&+& w (\Delta_{\rho_k^B} N_B)^2+ (1-w) (\Delta_{\rho_k^B} b)^2\nonumber\\
&+&wN_k^2 - w\exrho{N}^2\bigg\},
\eea

In Appendix A.3 we prove the following uncertainty relation
\be
R_{\rho,w}:=w(\Delta_{\rho} N_A)^2+(1-w)(\Delta_{\rho} a)^2 \ge
L_w(\exrho{N_A}),
\label{Rwrho}
\ee
where
\be
L_w(N)=\bigg\{ \begin{array}{ll} 
\sqrt{w(1-w)(N+\frac{1}{4})+\frac{w}{4}}-\frac{1}{2} &  \textrm{if $N\ge N_L$}, \\
(N-N_L)w(1-w)                       &  \textrm{if $N<N_L$}.
 \label{LwN}                                             
\end{array}
\ee
Here $N_L=(1-w)/4w$. Ineq. (\ref{Rwrho}) is the generalization of (\ref{Rrho}) for unequal weights
for the two variances.
For $N<N_L$ the function $L_w(N)$ is linear and the slope
is determined in such a way that there is not an abrupt change in the
derivative of $L_w(N)$ at $N=N_L$.

In order to get a lower bound for $\Sigma_{\rho,w}$ using the uncertainty
relation (\ref{Rwrho}), one has to follow similar steps as in Sec. II.A.
Using the facts that $L_w(N_1)+L_w(N_2)\ge L_w(N_1+N_2)+L_w(0)$ is fulfilled and $L_w(N)+wN^2$
is a concave function of $N$, the lower bound for $\Sigma_{\rho,w}$
is obtained as $f_w(N)=L_w(N)+L_w(0)$.

\subsection{Proof of main result}

\begin{figure}
\centerline{\epsfxsize=3.in\epsffile{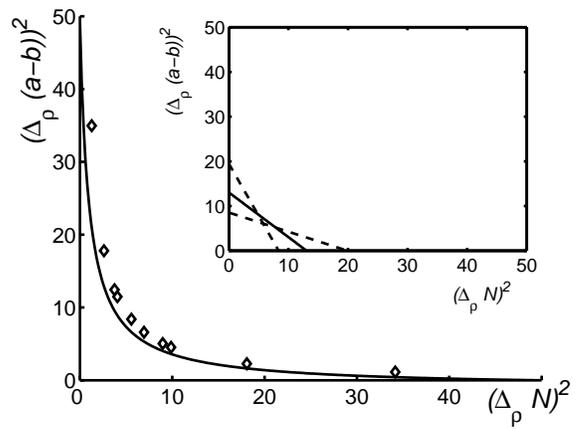}}
\caption{Numerical test of the inequality (\ref{vnvbregion}) for
the two-mode separability problem for N=200.
(solid) Boundary of the region defined by Ineq. (\ref{vnvbregion}).
All states below this line are entangled.
(diamonds) Points corresponding to separable states found numerically.
Inset: (solid) Boundary of the region defined by Ineq. (\ref{IneqNb});
(dashed) Boundary of the region defined
by Ineq. (\ref{Sigmaw}) for $w=0.3$ and $0.7$.}
\label{fig_region}
\end{figure}

In this subsection 
Ineq. (\ref{vnvbregion}) will be obtained by
determining the curve which has the lines corresponding to
different $w$'s as its tangentials.
The tangentials of a hyperbola $(y+c_0) = C/(x+c_0)$ are
given by $wx+(1-w)y=2\sqrt{w(1-w)C}-c_0$. One can reformulate
(\ref{Sigmaw}) by replacing the right hand side by
a slightly weaker lower bound which fits this form
\begin{equation}
\tilde{f}_w(N)=\sqrt{w(1-w)(N+\frac{1}{2})}-1.
\end{equation}
Hence the equation for a hyperbola
on the $\varho{N}$ -- $\varho{(a-b)}$ plane
can be obtained  (solid curve in Fig. \ref{fig_region}).
Ineq. (\ref{vnvbregion}) corresponds
to points above this hyperbola.
Any state which violates this inequality is
necessarily entangled. 

\section {Discussion}

First, the tightness of the necessary condition for separability 
(\ref{vnvbregion}) should be verified.
Numerical checks show
that it is quite strong (see Fig. \ref{fig_region}).
The diamonds indicate product states of the form
$\big|0\big> \otimes \big|\Psi\big>$ found numerically.
The state in the origin of Fig. \ref{fig_region},
giving zero for both variances in Ineq. (\ref{vnvbregion}),
is state (\ref{PPS}) as can be shown as follows. 
Eigenstates of
$\hat{N}$ with $N$ particles have the form
$|\Psi\rangle=\sum c_n|n,N-n\rangle$.
The state $(a-b)|\Psi\rangle$ has $N-1$ particles,
thus $(a-b)|\Psi\rangle=\lambda|\Psi\rangle$ is possible only if the
eigenvalue $\lambda=0$.
A state for which $(a-b)|\Psi\rangle=0$, has to fulfill
$c_{n+1} \sqrt{n+1}=c_n\sqrt{N-n}$. This determines the state
uniquely as (\ref{PPS}).

Our method detects entangled states in the proximity of state (\ref{PPS})
on the $\va{N}$ -- $\va{(a-b)}$ plane as shown in Fig. \ref{fig_region}.
(In this section we will omit the $\rho$ index.)
Other interesting states on the $\va{N}$ -- $\va{(a-b)}$ plane:
A separable state having $\va{N}=0$ is the convex
combination of products of Fock states
$\ket{n_k}\ket{N-n_k}$.
For these $\va{(a-b)}=N$.
Separable states with perfect destructive
interference between the modes
having $\va{(a-b)}=0$ are the convex combination
of products of coherent states of the form
$\big|\alpha_k\rangle\big|\alpha_k+c\rangle$ where
$c$ is a constant common for all product subensembles.
For these states  $\va{N} \ge N$ \cite{CS01}.

According to our criterion, to detect
entanglement in an experiment, the variances of
$N$  and $(a-b)$ should be measured. 
A simpler scenario is to measure the variance of $N$
and $\ex{(\ad-\bd)(a-b)}$.
Since $\va{(a-b)}\le\ex{(\ad-\bd)(a-b)}$ the (\ref{IneqNb})
inequality can be used
for this case after replacing $\va{(a-b)}$ by 
$\ex{(\ad-\bd)(a-b)}$.
The latter is just twice the particle number
in the mode $b'=(a-b)/\sqrt{2}$.

The generation of state (\ref{PPS}) is never perfect,
thus the system is in a mixed state. Our method makes it possible
to detect entanglement even in this case.
If $\langle b'^\dagger b'\rangle\approx\va{N}$
the maximum particle number variance for a state to be detected is
$\va{N}\propto\sqrt{N}$
which  is much smaller than for
coherent states. On the other hand, for perfect destructive 
interference when  $\langle b'^\dagger b'\rangle\approx 0$
the maximal variance is $\va{N}\propto N$.

Equ. (\ref{PPS}) describes the quantum state
of a Bose-Einstein condensate of atoms, if 
the $a$ and $b$ modes correspond to 
the two halves of the condensate \cite {CS01}.
In this case $(\ad+\bd)$ creates a particle in state $\big|\Psi\rangle$
and (\ref{PPS}) describes a product of single particle states
of the form
$|\Psi\rangle \otimes|\Psi\rangle\otimes\cdot\cdot\cdot\otimes |\Psi\rangle$.
Although it is a
product state from the point of view
of the individual particles, 
in the $\big|n,m\rangle$ basis it is clearly
entangled. 
In order to detect entanglement, 
one needs to measure the variance of the total particle
number and 
the particle number in one of the new modes after 
the two halves of the condensates interfere \cite{AK98,TS00}.

The condensate can be "split" into two modes, realizing 
state (\ref{PPS}), and then reunited for detection in a Mach-Zehnder 
type interferometer \cite{TS00}.
The state (\ref{PPS}) can also be obtained
in a Bose-Einstein condensate of two level atoms,
by preparing the atoms in the same
internal state and then applying a $\pi/2$ laser pulse.
The modes can then be spatially separated
with a state-dependent potential \cite{DS00}.

Finally, the state (\ref{PPS}) can be prepared
with a $50/50$ beam splitter and a
laser pulse corresponding to a state with low photon number variance.
For obtaining $\va{N}$ and $\langle b'^\dagger b'\rangle$,
a second beam splitter can be used, together with photon number measurements
in the two modes. In oder to detect entanglement, 
assuming perfect destructive interference at the second beam splitter,
for the photon source $\va{N} \le N/4-7/8$ is required.
This can be obtained, for example, with a state
 with sub-Poissonian number statistics.
            
Beside experimental considerations, the advantage of our approach is 
the ability to detect states in the vicinity of 
the entangled state (\ref{PPS}) 
which is not
detected by the method based on the correlation matrix \cite{DG00,RS00}.
The correlation matrix $\gamma$ contains the
correlations of two pairs of conjugate single-party observables,
which now we choose to be $\{R_k\}=\{x_A,p_A,x_B,p_B\}$.
Here $x_A=(a+\ad)/\sqrt{2}$, $p_A=(a-\ad)/(\sqrt{2}i)$, and 
$x_B$ and $p_B$ are defined similarly for the $b$ mode.
For the state (\ref{PPS}) the correlation matrix 
$\gamma_{kl}=Tr\{\rho (R_k-\langle R_k\rangle)
(R_l-\langle R_l\rangle)\}+Tr\{\rho
(R_l-\langle R_l\rangle) (R_k-\langle R_k \rangle)\}$ 
is obtained as
\begin{equation}
\gamma=\left( \begin{array}{cccc} 
N+1 & 0   & N   &  0  \\ 
0   & N+1 & 0   & N  \\
N   & 0   & N+1 &  0  \\
0   & N  & 0   & N+1 \\
\end{array}\right).
\label{corrmat}
\end{equation} 
The sufficient condition for inseparability is
$\tilde{\gamma} - iJ \ngeqslant 0$ where 
$\tilde{\gamma}$ is the correlation matrix corresponding to the partially
transposed density matrix and $J_{kl}=i[R_k,R_l]$.
Here $\tilde{\gamma} - iJ \geqslant 0$ thus
the state is not detected as entangled.

Moreover, with the simple method used for criterion (\ref{crit1}) described in the 
introduction, our criteria
(\ref{vnvbregion}), (\ref{IneqNb}) and (\ref{Sigmaw}) 
cannot be reduced to an entanglement witness.
This is because they contain the variance of the particle number and
$\ex{N}$ cannot be set to zero by single-party unitary
operations. 

\section {Conclusions}

In summary, a simple inequality for the expectation values of observables was
proposed for entanglement detection. Since only the measurement of
easily accessible quantities
(particle numbers and particle number variances) are needed,
this approach may be feasible for detecting entanglement
experimentally in Bose-Einstein condensates or in a two-mode photon field.

Our method can be generalized 
for detecting other highly entangled states.
First two operators must be identified which have the state
as an eigenstate. Then a necessary condition for separability
must be constructed with the variances of these operators.
Such a highly entangled state is for example the $\ket{N,0}+\ket{0,N}$
Schr\"odinger cat state which is the eigenstate of $N$ and 
$(\ad b)^N+(a\bd)^N$. 

\section {Acknowledgment}

G.T. would like to thank J.J. Garc\'{\i}a-Ripoll, B. Kraus and
 M.M. Wolf for useful discussions. G.T. and J.I.C. also acknowledge the
support of the EU project RESQ and QUPRODIS and the Kompetenznetzwerk
Quanteninformationsverarbeitung der Bayerischen Staatsregierung.
C.S. is supported by a Marie Curie Fellowship of the European
Union (HPMF-CT-2001-01205)

\appendix

\section{Single mode uncertainty relation}
\subsection{Analytic calculation}

In this subsection we will prove Ineq. (\ref{Rrho}). We will
find a lower bound for the sum of 
the two variances $(\Delta_{\rho} N_A)^2$
and $(\Delta_{\rho} a)^2$  for any
single-mode quantum state. This uncertainty relation
is needed to find a lower bound for the sum of
operator variances for two-mode separable 
states in Ineq. (\ref{IneqNb}).

The first term on the left hand side of Ineq. (\ref{Rrho}) is
zero for number states.
The second term is zero for coherent states. 
$N_A$ and $a$ have a common eigenvector: 
for the state $\ket{0}$ the variances of both are zero.
In order to find a non-trivial relation, 
the lower bound for the sum of the two variances
must have at least one parameter.
We choose this parameter to be $\exrho{N_A}$.
For $\exrho{N_A}>0$ the operators $N_A$ and $a$
do not have common eigenvectors and the sum of the two variances
are bounded from below. 

The proof of Ineq. (\ref{Rrho}) is based on finding two lower bounds
for the left hand side of Ineq. (\ref{Rrho}) and then combining them.
Let us denote $\langle a\rangle_\rho = \sqrt{\alpha \exrho{N_A}}e^{i\phi}$,
where $0\le \alpha\le 1$. The first lower bound is
\be
\label{x1}
R_\rho\ge\varho{a}=(1-\alpha)\exrho{N_A}=:B_1(N_A,\alpha).
\ee
The second bound \cite{NonHUnc} is obtained from
\bea
R_\rho&=&\varho{N_A}+\frac{1}{2}[\varho{x_A}+
\varho{p_A}]-\frac{1}{2}\nonumber\\
&\ge&\sqrt{2\varho{N_A}[\varho{x_A}+
\varho{p_A}]}-\frac{1}{2}.
\eea
Here for the inequality $X^2+Y^2 \ge 2XY$ was applied.
Now, using the facts that $\varho{N_A}\varho{x_A}\ge|\exrho{p_A}|^2/4$ and
$\varho{N_A}\varho{p_A}\ge|\exrho{x_A}|^2/4$ we obtain
\bea
\label{x2}
R_\rho  &\ge& \sqrt{ \frac {|\exrho{x_A}|^2+|\exrho{p_A}|^2}{2}}
-\frac{1}{2}=|\ex{a}|-\frac{1}{2}\nonumber\\
&=&\sqrt{\alpha \exrho{N_A}}-
\frac{1}{2}=:B_2(N_A,\alpha).
\eea
From Ineqs. (\ref{x1}) and (\ref{x2})
one can derive a higher lower bound for Ineq. (\ref{Rrho}) 
by taking the maximum of these two bounds. It can be shown that
\begin{equation}
B(N,\alpha):=\max \big[B_1,B_2\big] = \bigg\{ \begin{array}{ll}
B_1(N,\alpha)& \textrm{if $\alpha \ge\alpha_L$},\\
B_2(N,\alpha)& \textrm{if $\alpha \le \alpha_L$},\\
\end{array}
\end{equation}
where 
\begin{equation}
\sqrt{\alpha_L}=\sqrt{1+\frac{3}{4N}}-
\frac{1}{2\sqrt{N}}.
\end{equation}
Here $\alpha_L$ is always non-negative, however, for small particle numbers 
$N<1/4$ it is larger than $1$. 

The lower bound for Ineq. (\ref{Rrho}) will be constructed by 
minimizing $B(N,\alpha)$ with respect to $\alpha$.
After some algebra one obtains
\be
\min_\alpha B(N,\alpha)=\bigg\{ \begin{array}{ll} 
\sqrt{N+\frac{3}{4}}-1& \textrm{if $N >\frac{1}{4}$},\\
0                     & \textrm{if $N \le \frac{1}{4}$}.\\
\end{array}
\label{bbb}
\ee
As stated in Sec. II.A, in order to use this result in the 
two-mode separability problem the bound should fulfill two criteria
(i) $L(N)+N^2$ should be concave; (ii) $L(N_1)+L(N_2)\ge L(N_1+N_2)+L(0)$.
Equ. (\ref{bbb}) does not fulfill (ii), 
thus a weaker bound satisfying this condition
has to be chosen. $L(N)$, as defined in Equ. (\ref{LN}), is such a bound.
It coincides with the bound (\ref{bbb}) for $N \ge 1/4$ while for
$N<1/4$ it is negative.

\subsection{Numerical verification}

In this subsection we prove by numerical calculations
that Equ. (\ref{LN}) is a tight lower bound for Ineq. (\ref{Rrho}).
We will determine the state vector minimizing the left hand side of Ineq.
(\ref{Rrho}) and the corresponding minimum
with the constraint $\langle a^\dagger a\rangle=N_A$.

\begin{figure}
\centerline{\epsfxsize=\columnwidth \epsffile{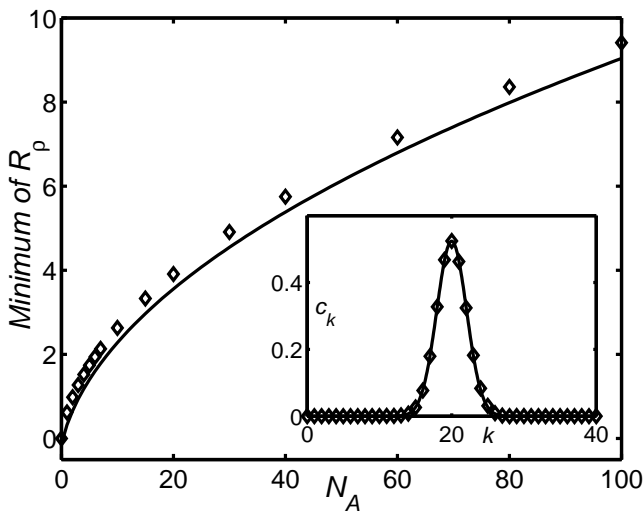} }
\vskip 0.2true cm
\caption{(diamonds) the minimum of $R_\rho$ (for a definition see
Equ. (\ref{Rrho}))
if the particle number is constrained to $N_A$.
(solid)  the analytic lower bound $L(N_A)$ defined in Equ. (\ref{LN}).
The inset shows the wave function in the number basis
(diamonds) corresponding to the minimum of uncertainties
for $N_A=20$. The results fit very well
a Gaussian curve (solid).}
\label{fig_min}
\end{figure}

The wave function is given in the number basis as
\begin{equation}
|\Psi\rangle = \sum_k c_k |k\rangle.
\end{equation}
The left hand side of Ineq. (\ref{Rrho}) can be rewritten as
\begin{eqnarray}
R_\rho&=& \bigg\{\sum_k |c_k|^2 k^2-N_A^2\bigg\}\nonumber\\
&+&\bigg\{N_A-\big|\sum_k c_k^*c_{k+1}\sqrt{k+1}\big|^2\bigg\}.
\end{eqnarray}
Lagrange multipliers must be added in order to
constrain the particle number to $N_A$ and keep the norm 1
\begin{eqnarray}
g(\{c_m\},\{c_m^*\},\mu_1,\mu_2) &=& R_\rho -\mu_1\big(N_A-\sum_k |c_k|^2 k\big)\nonumber\\
&-&\mu_2\big(1-\sum_k |c_k|^2\big).
\label{what_to_minimize}
\end{eqnarray}
When minimized, all the derivatives of the function 
(\ref{what_to_minimize}) 
must be zero. Moreover, since $R_\rho(\{c_k\}) \ge R_\rho(\{|c_k|\})$
we can restrict our search for the minimum for
real $c_k$'s. Hence one obtains
\begin{equation}
c_{n+1}=\bigg(\frac{n^2+\mu_1 n+\mu_2}
{A\sqrt{n+1}}\bigg)c_n-\bigg(\sqrt{\frac{n}{n+1}}\bigg)c_{n-1},
\label{series}
\end{equation}
where $A=\langle a\rangle=\sum c_k c_{k+1}\sqrt{k+1}$
and the term with $c_{n-1}$ is not present for $n=0$.
Based on Equ. (\ref{series}),
from $A$, $\mu_1$ and $\mu_2$ the unnormalized 
wave function can be constructed by setting $c_0=1$. 
As can be seen in Fig. \ref{fig_min}, 
$L(N_A)$ defined in Equ. (\ref{LN}) 
is very close to the minimum found numerically
thus it is a tight bound.
The wave function minimizing $R_\rho$
is shown in the inset. In the number basis it fits
very well a Gaussian curve even for small particle numbers.

\subsection{Generalized single mode uncertainty relation}

In this subsection we will prove Ineq. (\ref{Rwrho}).
For $w=0$ the state minimizing $R_{\rho,w}$ is a coherent state,
for $w=1$ it is a number state. For intermediate $w$'s the wave function
giving the minimum interpolates between these two.
The $L_w(\exrho{N_A})$ bound can be obtained, after inserting $w§$ and $(1-w)$
in the expression to be minimized, by following the same steps  as
in Appendix A.1. The two bounds found will be
\bea
B_{1,w}(N_A,\alpha)&=&(1-w)(1-\alpha)N_A,\nonumber\\
B_{2,w}(N_A,\alpha)&=&\sqrt{w(1-w)\alpha N_A}-\frac{1-w}{2}.
\eea
The maximum of these two, $B_w(N,\alpha)$, can be obtained knowing that
$B_{1,w}(N,\alpha)> B_{2,w}(N,\alpha)$ if $\alpha>\alpha_{L}$ where
\begin{equation}
\sqrt{\alpha_{L}}=\sqrt{\frac{2-w}{4N(1-w)}+1}-
\sqrt{\frac{w}{4N(1-w)}}.
\end{equation}
Hence the lower bound for $R_{\rho,w}$ is obtained as
\bea
\min_\alpha B_w=\bigg\{ \begin{array}{ll} 
\sqrt{w(1-w)(N+\frac{1}{4})+\frac{w}{4}}-\frac{1}{2}  
                      & \textrm{if $N >N_L$},\nonumber\\
0                     & \textrm{if $N \le N_L$},
\end{array}\\
\label{bbbb}
\eea
where $N_L=(1-w)/4w$. 

As stated in Sec. II.B, in order to use these results in the 
two-mode separability problem the bound should fulfill two criteria
(i) $L(N)+wN^2$ should be concave; (ii) $L(N_1)+L(N_2)\ge L(N_1+N_2)+L(0)$.
Equ. (\ref{bbbb}) does not fulfill (ii), 
thus a weaker bound satisfying both conditions
has to be chosen. $L_w(N)$, as defined in (\ref{LwN}) is such a bound.
It coincides with Equ. (\ref{bbbb}) for $N \ge N_L$ while for
$N<N_L$ it is a linear function of $N$ and it is negative.
The function giving $L_w(N)$
for $N>N_L$ (top line in Equ. (\ref{bbbb})) 
cannot simply be extended to $N\le N_L$ as it was done
in Appendix A.1 for the simpler uncertainty relation,
since in this case (i) would be not satisfied.



\begin{thebibliography}{99}
\bibitem{TermsKept} Notice, that we kept the non-negative term
$\sum_k N_k^2 - \exrho{N}^2$ in (\ref{summarho0}), 
which would be neglected if we followed the approach based
on uncertainty relations with constant bounds \cite{DG00,HT02}.
\bibitem{NonHUnc} The following deduction can be generalized for the sum 
of uncertainties of a Hermitian (X) and a non-Hermitian (Y=Q+iP) operator as
$\varho{X}+\varho{Y}\ge|\exrho{[X,Y]}|+i\exrho{[P,Q]}$. (Here $P$ and $Q$
are Hermitian operators.) If both $X$ and $Y$ are Hermitian 
(i.e.,$P=0$)
the inequality reduces to $\varho{X}+\varho{Y}\ge|\exrho{[X,Y]}|$. 
This latter expression can also be obtained from the usual 
product form of the Heisenberg uncertainty $\varho{X}\varho{Y}\ge |\exrho{[X,Y]}|^2/4$ 
using the inequality $\varho{X}+\varho{Y}\ge 2\sqrt{\varho{X}\varho{Y}}$.
\end{thebibliography}
\end{document}